\documentclass[12pt,preprint]{aastex}

\def\deg{\ifmmode^\circ\else$^\circ$\fi}

\def\mic{~$\mu$m}

\def\mic{$\mu${\rm m}}
\def\lir{{\rm L}$_{IR}$}

\def\arcs{\ifmmode {''}\else $''$\fi}
\def\arcm{\ifmmode {'}\else $'$\fi}
\def\parcs{\sa=.07em \sb=.03em
     \ifmmode $\rlap{.}$^{\scriptscriptstyle\prime\kern -\sb\prime}$\kern -\sa$
     \else \rlap{.}$^{\scriptscriptstyle\prime\kern -\sb\prime}$\kern -\sa\fi}
\def\parcm{\sa=.08em \sb=.03em
     \ifmmode $\rlap{.}\kern\sa$^{\scriptscriptstyle\prime}$\kern-\sb$
     \else \rlap{.}\kern\sa$^{\scriptscriptstyle\prime}$\kern-\sb\fi}

\def\Msun{M$_{\odot}$}
\def\Lsun{L$_{\odot}$}
\def\Myr{\Msun/yr}

\def\spose#1{\hbox to 0pt{#1\hss}}
\def\simlt{\mathrel{\spose{\lower 3pt\hbox{$\mathchar"218$}}
     \raise 2.0pt\hbox{$\mathchar"13C$}}}
\def\simgt{\mathrel{\spose{\lower 3pt\hbox{$\mathchar"218$}}
     \raise 2.0pt\hbox{$\mathchar"13E$}}}
\def\lsim{\rlap{$<$}{\lower 1.0ex\hbox{$\sim$}}}
\def\gsim{\rlap{$>$}{\lower 1.0ex\hbox{$\sim$}}}

\begin{document}

\title{Rest-Frame MIR Detection of an Extremely Luminous Lyman Break Galaxy with the Spitzer IRS\footnotemark\ \footnotemark
}

\footnotetext[1]{based on observations obtained with the Spitzer Space Telescope, which is 
operated by JPL, California Institute of Technology for the National Aeronautics and Space Administration}

\footnotetext[2]{The IRS is a collaborative venture between Cornell
 University and Ball Aerospace Corporation that was funded by NASA
  through JPL.}

\author{H. I. Teplitz\altaffilmark{3},
V. Charmandaris\altaffilmark{4,5}, 
L. Armus\altaffilmark{3}, 
P.N. Appleton\altaffilmark{3}, 
J.R. Houck\altaffilmark{4},
B.T. Soifer\altaffilmark{3}, 
D. Weedman\altaffilmark{4}, 
B.R. Brandl\altaffilmark{6},
J. van Cleve\altaffilmark{7}, 
C. Grillmair\altaffilmark{3},
K.I. Uchida\altaffilmark{4}
}

\altaffiltext{3}{Spitzer Science Center, MS 220-6, Caltech, Pasadena, CA 91125.  hit@ipac.caltech.edu}
\altaffiltext{4}{Astronomy Department, Cornell University, Ithaca, NY 14853}
\altaffiltext{5}{Chercheur Associ\'e, Observatoire de Paris, F-75014, Paris, France}
\altaffiltext{6}{Leiden University, 2300 RA Leiden, The Netherlands}
\altaffiltext{7}{Ball Aerospace and Technologies Corp. 1600 Commerce St., Boulder, CO 80301  }

\begin{abstract}
  
  We present the first rest-frame $\sim 4$\mic\ detection of a Lyman
  break galaxy.  The data were obtained using the 16\mic\ imaging
  capability of the Spitzer Infrared Spectrograph.  The target object,
  J134026.44+634433.2, is an extremely luminous Lyman break galaxy at z=2.79
  first identified in Sloan Digital Sky Survey spectra (Bentz, Osmer,
  \& Weinberg 2004).  The source is strongly detected with a flux of
  $0.94\pm 0.02$\ mJy.  Combining Spitzer and SDSS photometry with
  supporting ground-based J- and K-band data, we show that the
  spectral energy distribution is consistent with an actively 
  star-forming galaxy.  We also detect other objects
  in the Spitzer field of view, including a very red MIR source.  We
  find no evidence of a strong lens amongst the MIR sources.

\end{abstract}


\section{Introduction}

Two types of galaxies account for the vast majority of star formation
at $z>2$: the Lyman Break galaxies (LBGs; Steidel et al.\ 1996, 2003),
and the ultra/hyperluminous infrared galaxies detected at
submillimeter and millimeter wavelengths (Blain et al.\ 2002 and
references therein; Bertoldi et al.\ 2000).  As both selections
identify strong star formation, it is tempting to assume they are
sampling a common population.  However, current observations do not
show a large overlap between the two samples.  Only a small percentage
of LBGs are detected with SCUBA, even though their high star formation
rates predict they should be far-infrared (FIR) luminous (Chapman et
al.\ 2000; Peacock et al.\ 2000).  The reverse is true as well -- the
majority of ultraluminous infrared sources (ULIRGs; $L_{IR}>10^{12}\ 
L_{\odot}$) are so highly extincted that they would be missed in
rest-frame UV-selected surveys for high redshift galaxies like the
LBGs (Meurer et al.\ 1999; Goldader et al.\ 2002).  This difference
may imply that the two methods select galaxies at different stages of
evolution, or with intrinsically different physical characteristics.
The Spitzer Space Telescope gives us the first opportunity to study
both populations across a wide range of mid-infrared (MIR)
wavelengths.

Recently, six extremely bright LBGs were discovered in Sloan Digital
Sky Survey (SDSS) spectra (Bentz, Osmer, \& Weinberg 2004; hereafter
BOW).  The spectra of the six sources are typical of LBGs, with bright
rest-frame UV continua and clearly detected stellar and interstellar
absorption lines.  They appear 2-3 magnitudes brighter than the most
luminous LBGs in the Steidel et al.\ sample.  None of the six show
spectral features that would be expected for AGN, such as broad or
high ionization emission lines.  BOW find no morphological or
environmental indicators which suggest lensing.  By contrast, the only
comparably bright known LBG is MS1512-cB58 (Yee et al.\ 1996; hereafter
cB58); it is strongly lensed by a dense foreground cluster (Seitz et al.\ 
1998), and has an obvious arclike shape which is discernible from
the ground.

In this paper, we present the first Spitzer detection of an LBG at
16\mic\ with the Spitzer Infrared Spectrograph (IRS; Houck et al.\ 
2004) ``blue'' Peak Up filter which is centered at 16\mic.  ``blue''
peak up array.  One of the SDSS LBGs, J134026.44+634433.2 (hereafter
S-LBG-1), was selected as a Science Verification target for Spitzer.
The object has a spectroscopic redshift of z=2.79.  Throughout the
paper, we will adopt a flat, $\Lambda$-dominated universe ($H_0 = 70$
km s$^{-1}$\ Mpc$^{-1}$, $\Omega_M=0.7, \Omega_{\Lambda}=0.7$).

\section{Observations}

We obtained 16\mic\ imaging of S-LBG-1 on 14 November 2003 with the
IRS onboard the Spitzer Space Telescope.  Between exposures, we moved
the telescope in a four point dither pattern, with separations 
of 4\arcs.  At each point in the pattern, two nod positions were
observed, and two cycles were taken at each nod position.  A total of
16 exposures were taken, with integration times of one minute each.

The data were reduced using the IRS pipeline at the Spitzer Science
Center (see chapter 7 of the Spitzer Observer's
Manual\footnote{http://ssc.spitzer.caltech.edu/documents/som/}).
Individual frames were registered based on the reconstructed pointing
(accurate to better than 1\arcs) and refined with centroiding of
objects.  Bad pixels were masked before mosaicing. We have not removed
the small ($< 2$\%) geometric distortion in the images.  We have
several times redundancy for each pixel on the sky, and we remove
faint latent images caused by bright sources in PU observations prior
to our program.

Photometric calibration utilized routine observations of a bright
standard.  We measured the flux in 16\mic\ images of HD 46190 to
calculate the zeropoint by comparison with template spectra based on
normalized Kurucz (1993) models (Cohen et al.\ 2003).  A curve of
growth was used to estimate total flux for a point source from
aperture photometry.

We also obtained supporting near-infrared (NIR) data in the J- and
K-bands with the WIRC instrument (Wilson et al.\ 2003) on the Hale
200-inch telescope.  The observations were taken on 27 January 2004.
The field was observed for five dithered exposures in each filter.
J-band exposures were two coadds of one minute each, and K-band were
two coadds of 30 seconds each.  The night was not photometric, but
conditions were stable over the few minutes of the observations.  The
WIRC field of view contains $\sim 20$\ 2MASS stars, which we used to
calibrate the photometry and estimate the extinction.

\section{Results}

The final, registered 16\mic\ image (Figure 1) has an RMS noise of 15\ $\mu$Jy
in an aperture with 5\arcs\ radius, and 25\ $\mu$Jy in an aperture of
9\arcs\ radius.  
S-LBG-1 is detected, with a flux of $0.94\pm 0.02$\ mJy.  Seven
other objects are  detected ($5\sigma$\ or more) in
the field, one of which is very bright in the MIR and quite faint at
optical wavelengths.  Of these, two have no counterpart in the SDSS or
WIRC imaging.  Five additional sources are marginally detected
($2-3~\sigma$), without optical counterparts.  However, given the
faintness of these sources  it is not surprising
that deeper optical/NIR data will be needed to identify them.
The reddest object, J134025.02+634358.3 (hereafter RED-1), is discussed
further in section \ref{sec: ero}.  Photometric data from IRS and WIRC
are given in Table 1.  The WIRC imaging shows no objects in the field
that were not detected in SDSS.

There is no indication that S-LBG-1 is resolved, but it is not
expected to be.  The 16\mic\ imaging is diffraction limited, and the
point spread function (PSF) has a full width at half maximum (FWHM) of
4\arcs.  S-LBG-1 is compact in both the SDSS and WIRC imaging, which
have FWHM of $\sim 0.75$\ and $\sim 1$\arcs, respectively.  At the
redshift of S-LBG-1, z=2.79, one arcsecond corresponds to $\sim 8$\ 
kpc.

\section{Discussion}

The 16\mic\ filter extends from 13.3 to 18.6 \mic\ at FWHM, or 3.5 to
4.9 \mic\ in the rest frame.  This wavelength range encompasses
Br$\alpha$\ but does not include the  polycyclic aromatic
hydrocarbon (PAH) feature centered at 3.3 \mic\ (Tokunaga et al.\ 
1991).  The filter samples the Rayleigh-Jeans tail of the stellar
photospheric continuum emission emission, which can be considerable if
the galaxy contains a substantial population of old stars.
Alternatively, 4 \mic\ marks the beginning of the rise in emission
from warm dust, heated either by very young stars or by an active
nucleus.

The spectral energy distribution (SED) of S-LBG-1 is consistent with a
strongly starbursting galaxy (Figure 2).  In the figure we show a
direct comparison with the spectrum of NGC 5253 (Wu et al.\ 2002), a
``benchmark starburst'' (Calzetti et al.\ 1999) with regions that have
undergone intense star formation in the last 100 Myr (see Calzetti et
al.\ 1997 and the reference therein), and a substantial older stellar
population.  It has MIR emission from dust heated by massive,
young stars (Crowther et al.\ 1999).  The SED of S-LBG-1 is redder
than NGC 5253 in the UV, but the two match quite well if one magnitude
of additional extinction is applied.  The rest-frame optical color of
S-LBG-1 (the observed J- and K-band) is bluer than that of NGC
5253, perhaps implying a lower fraction of old stars.

Ellingson et al.\ (1996) showed that the optical/NIR SED of cB58 could
be fit with a recent (10 Myr) secondary burst of star-formation added
to a large (85\% by mass) stellar population with an age of $\lsim 1$\ 
Gyr.  The young age of the source was later confirmed by UV absorption
line studies (Pettini et al.\ 2002).  A marginal detection of cB58
with ISOCAM (Bechtold et al.\ 1998), at rest-frame 1.8 and 3 \mic,
favored a higher percentage of younger stars.  Taken at face value,
the cB58 SED is not consistent with S-LBG-1 unless cB58 rises steeply
beyond the rest-frame K-band.  The inconsistency could result from a
higher dust content in S-LBG-1 and likely a dissimilar spatial
distribution of dust, which may not be surprising given the different
star formation states of the sources.  The intrinsic (corrected for
lensing magnification) SFR of cB58 is $\sim 20$\ \Myr\ (Teplitz et
al.\ 2000), compared to the $\sim 1000$\ \Myr\ in S-LBG-1 (without
correcting for extinction).

\subsection{Possible AGN Contribution}

The S-LBG-1 SED is also consistent with a relatively dust-free Seyfert
galaxy (Figure 2).  However, BOW see no high excitation lines in the
UV.  It is also possible that the observed 16\mic\ flux of S-LBG-1
indicates a ``buried'' AGN that is not seen in the UV.  BOW discuss
the possibility that the S-LBG's are BAL QSO's, but conclude that they
would be unique amongst that class of objects if they were.
Nonetheless, the luminosity of S-LBG-1 may favor the possibility of
AGN activity dominating the MIR flux.  In addition, shorter wavelength
Spitzer observations of S-LBG-1 are needed to conclusively rule out a
very steep MIR slope, given the low rest-frame 3 \mic\ flux of cB58.

Laurent et al. (2000) show that the presence of an AGN is revealed in
the MIR by an excess of emission in the 3-6 \mic\ range.  This has
been attributed to hot dust emission heated to near sublimation
temperatures ($\sim 1000$\ K for silicates and $\sim 1500$\ K for
graphite) by the accretion disk. The presence of this continuum has
been detected in a number of nearby galaxies hosting an active AGN
including NGC1068 and NGC4151 (Le Floc'h et al. 2000, Alonso-Hererro
et al. 2003), as well as Centaurus A (Mirabel et al. 1999). However,
such excess usually has a positive slope in the 3-6 \mic\ range,
making it challenging to identify in distant systems because the old
stellar population of the bulge of the galaxy may be missinterpreted
as an excess of thermal emission.  Furthermore, there has been
evidence that as the luminosity of dust enshrouded IR galaxies
increases beyond $10^{12.3}$\ \Lsun, so does the probability that an
AGN contributes substantially to the heating of the dust (Sanders et
al.  1988; Veilleux, Kim, \& Sanders 1999; Tran et al.\ 2001).

If the brightest LBGs harbor buried AGN, it might indicate that the
typical percentage of AGN in LBG searches, $\sim 10$\% (Steidel et
al.\ 2002), is underestimated.  In that case, the inferred \lir\ 
of LBG systems, and thus the large correction to their inferred contribution to 
the global density of star formation, would be overestimated.

\subsection{Comparison with Ultraluminious IR sources}

The prodigious star formation rates of S-LBG-1 ($\gsim 1000$\ \Myr;
BOW) implies that it must be generating enough UV radiation to put it
in the same luminosity class as ULIRGs ($L_{IR} > 10^{12}\ L_{\odot}$)
and perhaps enough to be comparable to the hyper-LIRGs found among the
SCUBA galaxies ($L > 10^{13}\ L_{\odot}$; Blain et al.\ 2002).  It is
not certain, however, how much of this radiation is absorbed and
reradiated by dust.  The optical to mid-infrared SED of S-LBG-1 is not
consistent with Arp 220, given the latter's high extinction in the UV.
However, the SED over our sampled wavelength provides only a tenuous
indication of the bolometric luminosity.

If the S-LBGs in general are not fundamentally different than ULIRGs,
then one must ask why isn't their large UV luminosity absorbed and
reradiated by dust.  LBGs are found to be 5-20 times underluminous
(less dusty) in the IR for their UV slopes (Baker et al.\ 2001; van
der Werf et al.\ 2001), while ULIRGs are equally overluminous
(Goldader et al.\ 2002).  One possibility is that they are at a
different stage of evolution.  Star formation in $z<1$\ ULIRGs is
triggered by merger activity (Flores et al.\ 1999).  In such systems,
the dust is dynamically mixed throughout and can easily absorb most of
the UV radiation emitted from actively star-forming HII regions.  If
the LBGs have a less homogeneous distribution of dust, more UV
radiation may escape.  Another possibility is that the extreme UV
radiation field of the most luminous LBGs heats the dust to higher
temperatures than in typical starbursts, suppressing the FIR emission
relative to the UV (Baker et al.\ 2001).  This would have the effect
of shifting the peak of the reradiated luminosity closer to the
Spitzer wavelengths.  Finally, a significantly lower dust content --
either the result of extremely low metallicity, or the destruction of
grains by the UV radiation field (as suggested by BOW) -- would allow
greater UV brightness.

\subsection{Lensing and the extremely red object}

\label{sec: ero}

In the SDSS image, BOW find no evidence to suggest strong lensing.
One might expect that if the lens were highly extincted, it could be
detected in the near- or mid-IR.  However, we find no evidence of a
massive source close in projection to S-LBG-1 that could be identified as a
gravitational lens.  The 16\mic\ PSF does not allow us to identify
close ($\lsim 3$\arcs) companions, but such objects would likely have
been seen in the K-band if they were present.    The critical radius is a
few arcseconds even for a massive elliptical (Blandford \& Narayan
1992; Eisenhardt et al.\ 1996).  The brightest MIR 
source in the field, RED-1, lies 30\arcs\ away.  

RED-1 is somewhat unusual in both brightness and color. It has a flux
density of 3 mJy at 16\mic, but is faint in the optical ($r_{AB}=23.3$). 
The surface density of such sources is
quite low.  ISO observations of the ELAIS field detected $14\pm 3$\ 
sources down to 3 mJy per square degree, or one per $\sim 200$\ PU
fields (La Franca et al.\ 2004).  Down to 1 mJy, the source counts are
only ten times higher (Elbaz et al.\ 1999 and the reference therein).
Few 3 mJy sources are as red as RED-1.  La Franca et al.\ 
find only 18\%\ of 15 \mic\ sources brighter than 1 mJy have $R$\ 
magnitudes fainter than 23.

The morphology of RED-1 is hard to quantify.  It is faint and has low
surface brightness in both the optical and NIR.  It is detected only
at the limit of the WIRC and SDSS images.  It shows no central central
concentration and is clearly extended, covering several square
arcseconds.

The 16 \mic\ filter samples the $\sim 12$\ \mic\ PAH feature and very
small grain continuum emission at $z\sim 0.3$\ and the $\sim 7$\ \mic\ 
PAH features at $z\sim 1$.  At the higher redshift, RED-1 would fall
into the hyper-LIRG luminosity class, but at $z\sim 0.3$\ it would be
only $10^{11}$\ \Lsun\ (Chary \& Elbaz 2001).  In addition, the object is
not extremely red in R-K as would be expected for a $z\sim 1$\ ULIRG.

\section{Summary}

We have detected an extremely luminous Lyman break galaxy in the first
Spitzer 16\mic\ image of a UV-selected source at high redshift.  The
MIR data point is consistent with the large star formation rate
inferred from the UV continuum.  The SED confirms that the star
formation is likely not extincted enough for the object to be
considered a ULIRG.  We find no evidence for a strong lens in the field.

Future Spitzer observations will be crucial to understanding the
connection between vigorously star forming LBGs and ULIRGs.  Deep
imaging with IRAC (Fazio et al.\ 2004) and MIPS (Rieke et al.\ 2004)
will potentially detect moderate luminosities LBGs.  The 16\mic\ 
imaging capability of the IRS is a powerful additional mode for the
study of these objects.  The detection of MIR emission in
S-LBG-1 demonstrates that the brightest LBGs will be observable by IRS
spectroscopy, allowing more detailed studies of their dust properties.
Such observations will open a new window onto a class of objects that
may be different from the infrared luminous sources that will be more
commonly studied with Spitzer.

\acknowledgements

We thank L. Yan, D. Frayer, and J. Colbert for helpful suggestions.
This work is based in part on observations 
made with the Spitzer Space Telescope, which is operated by the Jet 
Propulsion Laboratory, California Institute of Technology under NASA 
contract 1407. Support for this work was provided by NASA through an 
award issued by JPL/Caltech.

\references

\reference{} Adelberger, K.L. \& Steidel, C.C.\ 2000, ApJ, 544, 218

\reference{} Alonso-Herrero, A.,  Quillen, A. C., Rieke, G. H., Ivanov, V. D., 
Efstathiou, A., 2003, AJ, 126, 81

\reference{} Baker, A. J., Lutz, D., Genzel, R., Tacconi, L. J., \& Lehnert, M. D. 2001, A\&A, 372, 37

\reference{} Bechtold, J., Elston, R., Yee, H. K. C., Ellingson, E., \&
 Cutri, R. M.\ 1998, in ``The Young Universe: Galaxy Formation and Evolution at Intermediate and High Redshift''. Eds
 S. D'Odorico, A. Fontana, and E. Giallongo. ASP Conference Series; Vol. 146, p.241

\reference{} Bentz, M.C., Osmer, P.S., \& Weinberg, D.H.\ 2004, ApJL, 600, 19 (BOW)

\reference{} Bertoldi, F., et al.\ 2000, A\&A,360,92

\reference{} Blain, A. W., Smail, I., Ivison, R. J., Kneib, J.-P., Frayer, D. T.\  2002, PhR, 369, 111

\reference{} Blandford, R. D. \& Narayan, R.\ 1992, ARA\&A, 30, 311

\reference{} Bruzual, G. \& Charlot, S.\ 2003, MNRAS, 344, 1000

\reference{} Calzetti, D., Conselice, C.J., Gallagher, J.S., III, Kinney, A.,L.\ 1999, AJ, 118, 797

\reference{} Chapman, S.C., et al.\ 2000, MNRAS, 319, 318

\reference{} Chary, R. \& Elbaz, D.\ 2001, ApJ, 556, 562

\reference{} Cohen, M., Megeath, T.G., Hammersley, P.L., Martin-Luis, F., \&
Stauffer, J. 2003, \aj, 125, 2645

\reference{} Crowther, P.A., Beck, S.C., Willis, A.J., Conti, P.S., Morris, P.W., Sutherland, R.S.\ 1999, MNRAS, 304, 654

\reference{} Edelson, R. A. \& Malkan, M. A.\ 1986, ApJ, 308, 59

\reference{} Eisenhardt, P.R., Armus, L., Hogg, D.W., Soifer, B.T., Neugebauer, G., \& 
Werner, Michael W.\ 1996, ApJ, 461, 72

\reference{} Elbaz, D., et al.\ 1999, A\&A, 351L, 37

\reference{} Ellingson, E., Yee, H.K.C,
        Bechtold, J., \& Elston, R. 1996, ApJL 466, 71

\reference{} Fazio, G., et al.\ 2004, ApJS, this volume

\reference{} Flores, H., et al.\ 1999, ApJ, 517, 148

\reference{} Goldader, J.D., et al.\ 2002, ApJ, 568, 651

\reference{} Houck, J. et al.\ 2004, ApJS, this volume

\reference{} Krolik, J. H., Horne, K., Kallman, T. R., Malkan, M. A., Edelson, R. A., \& Kriss, G. A.\ 1991, ApJ, 371, 541

\reference{} Kurucz, R. L. 1993, CD-ROM 13, ATLAS9 Stellar Atmosphere 
 Programs and 2 km/s Grid (Cambridge: Smithsonian Astrophys. Obs.)

\reference{} La Franca, F., et al.\ 2004, AJ, in press; astro-ph/0403211

\reference{} Laurent, O., Mirabel, I.F., Charmandaris, V., Gallais, P., Madden, 
S.C, Sauvage, M., Vigroux, L., \&  Cesarsky, C. 2000, A\&A, 359, 887

\reference{} Le Floc'h, E.,  Mirabel, I.F., Laurent, O., Charmandaris, V., Gallais, 
P., Sauvage, M., Vigroux, L., \&  Cesarsky, C.  2001, A\&A, 367, 487

\reference{} Meurer, G.R., Heckman, T.M., \& Calzetti, D.\ 1999, ApJ, 521, 64

\reference{} Mirabel, I.F. et al. 1999, A\&A, 341, 667

\reference{} Peacock, J.A., et al.\ 2000, MNRAS, 318, 535

\reference{} Pettini, M., Rix, S.A., Steidel, C.C., Adelberger, K.L., Hunt, M.P., \& Shapley, A.E.\ 2002, ApJ, 569, 742

\reference{} Rieke, G. et al.\ 2004, ApJS, this volume

\reference{} Sanders, D. B., Soifer, B. T.,  Elias, J. H.,  Madore, B. F., 
Matthews, K., Neugebauer, G., \& Scoville, N. Z., 1988, ApJ, 325, 74

\reference{} Seitz, S., Saglia, R.P., Bender, R., Hopp, U., Belloni,
P., \& Ziegler, B.  1998, MNRAS 298, 945

\reference{} Shapley, A.E., Steidel, C.C., Pettini, M., \& Adelberger, K.L.\ 2003, ApJ, 588, 65

\reference{} Silva,Ll, Granato, G.L., Bressan, A., \& Danese, L.\ 1998, ApJ, 509, 103

\reference{} Steidel, C. C., Adelberger, K. L., Shapley, A. E., Pettini, M.,Dickinson, M., \& Giavalisco, M.\ 2003, ApJ, 592, 
728

\reference{} Steidel, C.C., et al.\  2002, ApJ, 576, 653

\reference{} Steidel, C.C., Adelberger, K.L., Giavalisco, M., Dickinson, M. \& Pettini, M.\ 1999, ApJ, 519, 1

\reference{} Teplitz, H.I. et al.\ 2000b, ApJLetters, 533, 65

\reference{} Tokunaga, A.~T.,
Sellgren, K., Smith, R.~G., Nagata, T., Sakata, A., \& Nakada, Y.\ 1991,
ApJ, 380, 452

\reference{} Tran, Q.D., et al.\ 2001, ApJ, 522, 527

\reference{} van der Werf, P.P., Knudsen, K.K., Labbe, I., \& Franx, M.\ 2001, astro-ph/0011217

\reference{} Veilleux, S., Kim, D.-C., Sanders, D.B.\ 1999, ApJ, 522, 113

\reference{} Yee, H.K.C., Ellingson, E., Bechtold, R.G., Carlberg,R.G., Cuillandre, J.-C. 1996, AJ, 111, 1783

\reference{} Wilson, J.C., et al.\ 2003, SPIE, 4841, 451

\reference{} Wu, W., Clayton, G.C., Gordon, K.D., Misselt, K.A., Smith, T.L., Calzetti, D.\ 2002, ApJS, 143, 377

\clearpage

\clearpage

\begin{deluxetable}{crrrrrrrr}
\tabletypesize{\scriptsize}
\tablecaption{Photometry\label{tbl-1}}
\tablehead{
\colhead{Object} &
\colhead{$u$} &
\colhead{$g$} &
\colhead{$r$} &
\colhead{$i$} &
\colhead{$z$} &
\colhead{$J$} &
\colhead{$K$} &
\colhead{16\mic} \\
\colhead{} &
\colhead{AB mag.} &
\colhead{AB mag.} &
\colhead{AB mag.} &
\colhead{AB mag.} &
\colhead{AB mag.} &
\colhead{AB mag.} &
\colhead{AB mag.} &
\colhead{mJy} 
}

\startdata
S-LBG-1 & 22.96 & 20.51 & 19.82 & 19.35 & 19.01 & 18.7$\pm 0.1$ & 18.8$\pm 0.1$ &  0.94$\pm 0.02$  \\
Red-1   & \nodata & 23.0$\pm 0.4$ & 23.3$\pm 0.6$ & 21.3$\pm 0.3$ & \nodata & 20.5$\pm 0.2$ & 20.0$\pm 0.2$ & 3.59$\pm 0.07$  

 \enddata
\tablecomments{S-LBG-1 optical magnitudes are from BOW.}

\end{deluxetable}

\clearpage
\begin{figure}[t*]

\hspace*{-.01in}

\plottwo{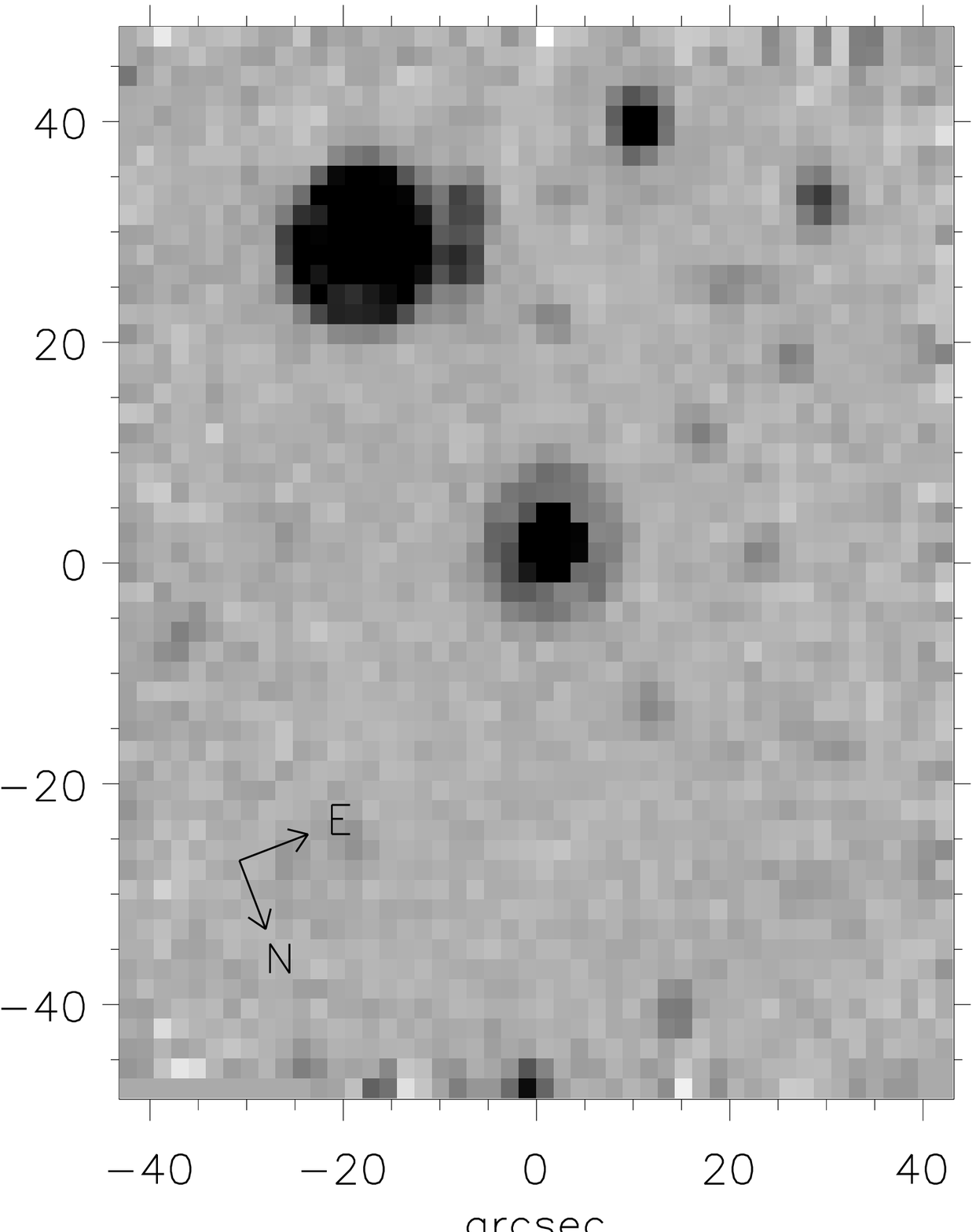}{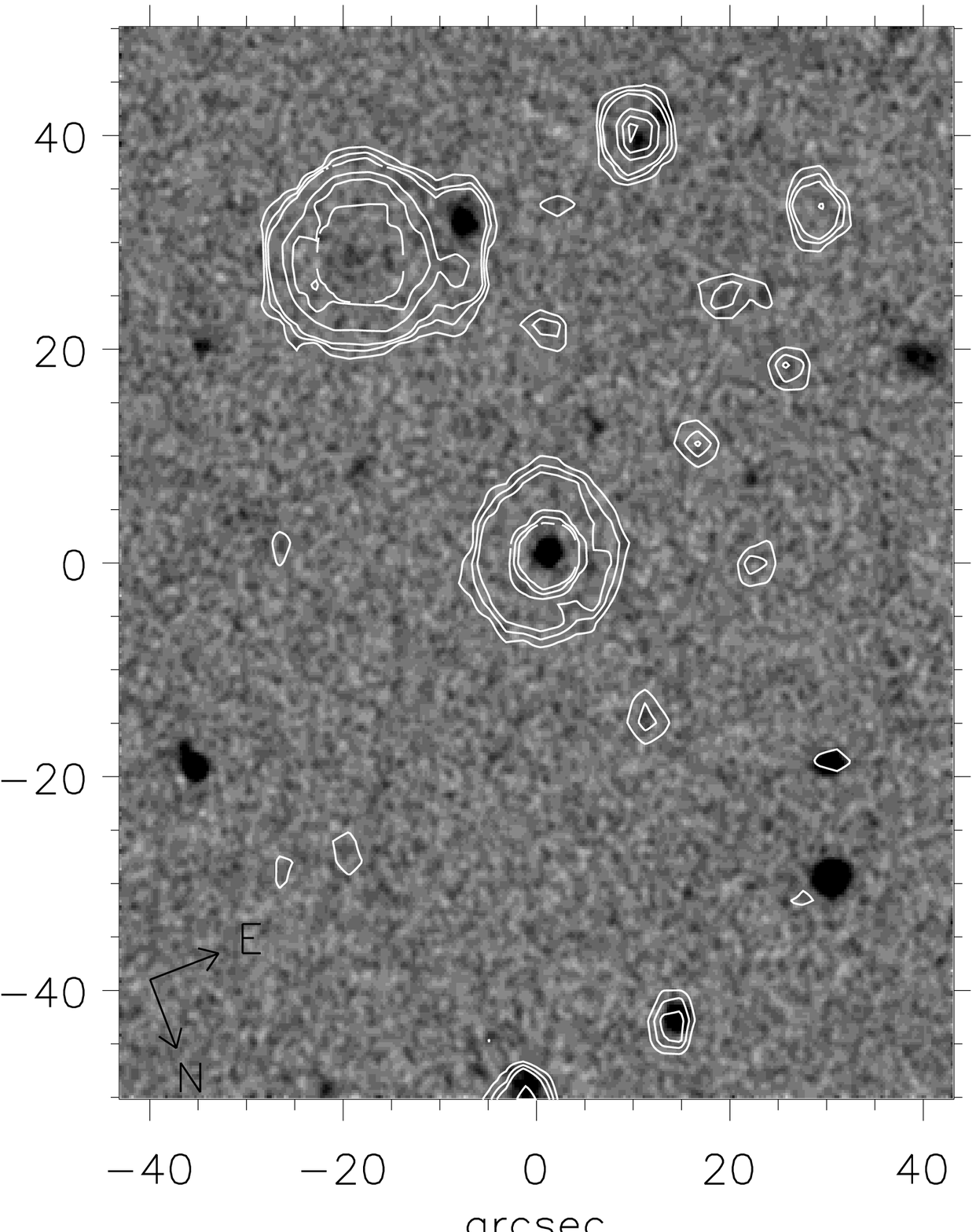}

\caption{(left) IRS 16\mic\ Spitzer image of the field.  S-LBG-1 
  is the object at the center.  Spatial offsets are indicated relative
  to S-LBG-1.  (right) Contours of detected and marginal objects in
  the Spitzer data overplotted on the SDSS $i$-band image.  The SDSS
  image has been rotated to match the orientation of the Spitzer
  image, and smoothed by the FWHM of its PSF.  Contours
  were plotted after scaling the Spitzer image to the SDSS pixel
  scale, with bilinear interpolation.  }

\end{figure}

\clearpage

\begin{figure}[t*]

\plotone{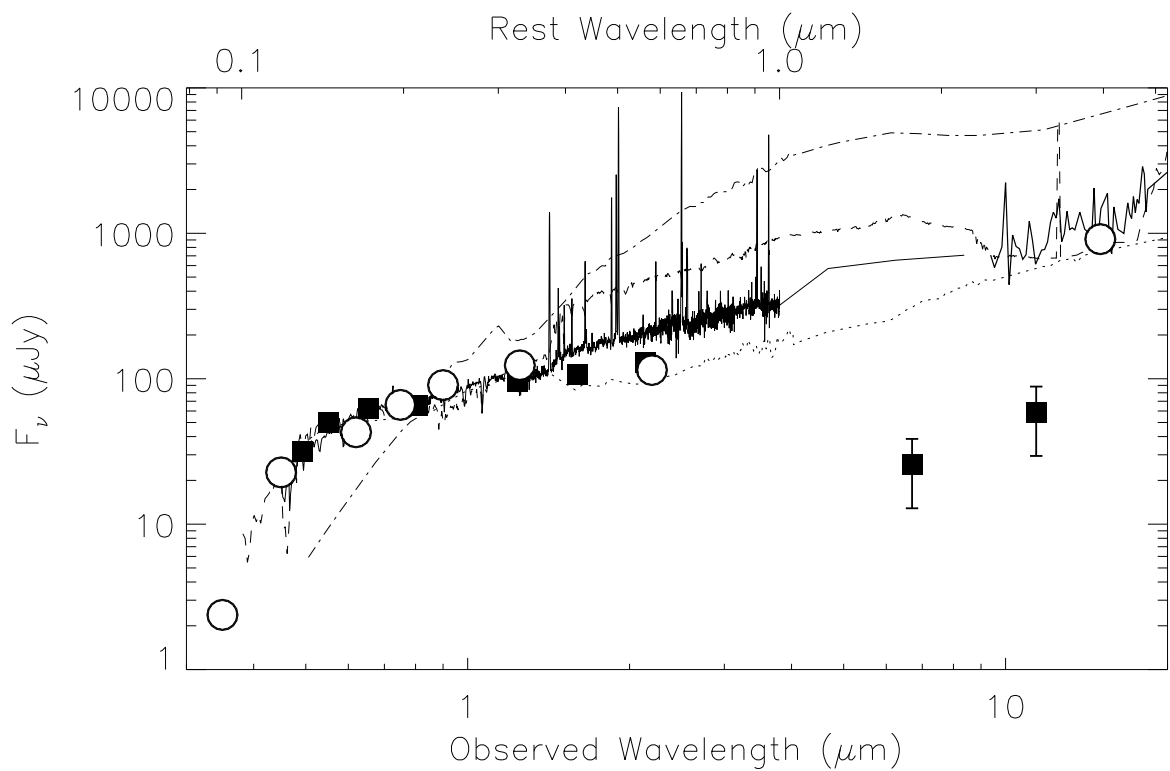}
\caption{
  We show the observed optical/IR SED of S-LBG-1 (open circles) at 0.3
  to 16 \mic.  Error bars are comparable to or smaller than symbol
  size.  We plot for comparison the spectral templates for an
  ultraluminous starburst galaxy (Arp 220; dashed line; Silva et al.\ 
  1998), two Seyfert 1s (NGC 3227, dot-dash line; NGC 5548, dotted
  line; Edelson \& Malkan 1986; Krolik et al.\ 1991), and the
  starburst galaxy NGC 5253 reddened by an additional $A_v=1$\ (solid
  line; Wu et al.\ 2002 and the references therein; ).  We also
  compare the SEDs of cB58 (filled squares; Ellingson et al.\ 1996 and
  Bechtold et al.\ 1999) The spectral templates and cB58 photometry
  have been scaled to match S-LBG-1 in the rest-frame UV.  }

\end{figure}

\end{document}